# Probing Internal Dynamics of Spatiotemporal Optical Vortex Strings: Spatiotemporal Attraction and Filament Stretching

Xiuyu Yao[1,2,*] †, Xuechen Gao[3]†, Ping Zhu[1,*], Jintao Fan[3,*], Jingwen Ran[4], Zezhao Gong[1,2], Dongjun Zhang[1], Xiao Liang[1], Xuejie Zhang[1], Meizhi Sun[1], Qiang Zhang[1,2], Lijie Cui[1], Hailun Zeng[1] , Minglie Hu[3], Xinglong Xie[1,*] and Jianqiang Zhu[1]

[1]Key Laboratory of High Power Laser and Physics, Shanghai Institute of Optics and Fine Mechanics, Chinese Academy of Sciences, Shanghai 201800, China.
[2]Center of Materials Science and Optoelectronics Engineering, University of Chinese Academy of Sciences, Beijing 100049, China.
[3]Tianjin University, School of Precision Instruments and Opto-Electronics Engineering, Key Laboratory of Opto-electronic Information Science and Technology of Ministry of Education, Ultrafast Laser Laboratory, Tianjin, 300072, China.
[4]Engineering Research Center of Mechanical Testing Technology and Equipment (Ministry of Education), Chongqing University of Technology, Chongqing 400054, China.
These authors contributed equally: Xiuyu Yao, Xuechen Gao.
* Corresponding authors: zhp1990@siom.ac.cn (P. Zhu); fanjintao@tju.edu.cn (J. Fan); yaoxy@siom.ac.cn (X. Yao); xiexl329@siom.ac.cn (X. Xie); jqzhu@siom.ac.cn (J. Zhu)

**Abstract:** Vortex dynamics are intriguing and challenging across multiple physics fields. In optics, customized spatiotemporally structured optical fields, especially spatiotemporal optical vortices (STOV), offer the potential to tailor light via coupled space-time degrees of freedom. However, the interaction mechanisms between multiple transverse orbital angular momentum singularities within a single wave packet remain elusive. This study explores the intrinsic dynamics of a STOV with three phase singularities, observing a pronounced vortex singularity oscillation phenomena by tuning the temporal dispersion. We show that these phenomena originate from the counterintuitive spatiotemporal attractive effect between vortices, which is closely related to the singularity distance. Furthermore, the stretching into filaments and annihilation behaviors is observed by introducing antivortex in the center of the wavepacket. Experimentally, we propose a Full Interferometric Retrieval of Spatiotemporal Tomography (FIRST) method that enables the complete, single-shot capture of wave packets, with excellent agreement between theoretical predictions and experimental results. To the best of our knowledge, the dynamics of transverse spatiotemporal singularities within a single wave packet are reported here for the first time. These findings confirm the existence of interesting interactions between STOV singularities, deepen our understanding of photonics and open a new direction for investigating the complex dynamics of vortex singularities in the spatiotemporal domain.

## Introduction

Recent progress in quantum physics and optics have pushed the boundaries of our understanding of spacetime, a concept that was revolutionized when Einstein's theory of relativity challenged Newton's notion of absolute spacetime. A prominent example is the theoretical prediction and experimental realization of time crystals, extending the concept from the spatial to the spatiotemporal (ST) domain (*1, 2*). Meanwhile, ST optical beams incorporate a temporal dimension, this endows ST beams with unprecedented properties, including arbitrary group velocities (*3*), anomalous refraction (*4*), and most notably—the generation of spatiotemporal optical vortices (STOV) (*5-7*). These vortices appear as a rotating phase in the spatiotemporal domains $(x, \tau)$, where the direction of the angular momentum is perpendicular to the wavevector. Such spatiotemporal vortex fields have been theoretically predicted (*5, 8*) and experimentally generated using linear devices (*6, 7*), attracting extensive attention and growing research interest (*9*).

Particularly, STOVs feature singularities as a key attribute of their topological structure, indicating the flow direction of the field and reveals the field's topology from an alternative perspective. These unique features hold great potential in high-dimensional information encoding and optical communication, optical metrology, quantum optics, strong-field physics, and other fields (*10-19*). Beyond isolated vortices, the coexistence of multiple singularities introduces the possibility of rich interaction dynamics. Since Indebetouw discovered the dynamical behavior of spatial vortices in 1993, various characteristics have been explored (*20-22*), such as vortex collisions (*23*), vortex knots (*24*), and in various optical systems (*25*). Vortices that vary with time are referred to as time-varying singularities. Understanding the dynamics and mutual interactions of vortices carrying transverse orbital angular momentum (OAM) is key to clarifying and taming high dimensional light. These have shown distinct application prospects and research appeal in light-matter interaction, optical communication, and other related fields. (*10, 12, 15, 17, 26, 27*).

However, previous studies have mainly focused on temporally independent wave packets (*10, 15*), leaving the interaction dynamics of multiple transverse spatiotemporal singularities within a single ultrafast wave packet largely unexplored. This limits the information density and the application prospects in ultrafast optical physics. Experimentally, accessing such dynamics remains challenging, as the commonly used time-domain scanning interferometry lacks sufficient temporal resolution and scanning stability imposed by the pulse width limitation of its reference arm.

Here, we demonstrate intriguing dynamics between phase vortex-vortex and vortex-antivortex strings within a single spatiotemporal wavepacket. We observe a "vortex dance" like phenomenon characterized by behaviors such as mutual attraction, stretch into filaments and annihilation of vortices. This novel optical process is validated both theoretically and experimentally: our experiments employ single-frame 3D imaging method to precisely track the trajectories of vortices in the optical field, confirming the existence of these intriguing features. Our findings reveal that vortex-vortex exhibit counterintuitive spatiotemporal attractive dynamics, while vortices-antivortices stretch each other, forming filaments, accompanied by mutual annihilation. Meanwhile, the observed spatiotemporal attraction and related phenomena can be controlled by the initial distance between vortices, which determines the strength of interactions between spatiotemporal vortices. To the best of our knowledge, the interaction dynamics of transverse spatiotemporal vortex singularities within a single wave packet are reported for the first time. This finding confirms the existence of interesting interactions between STOV singularities, advances the in-depth understanding of vortex interactions in fields such as optical physics and fluids but also open up a new direction for extending the study of complex electrodynamics into the spatiotemporal domain.

## Results

### *Attractive effect of STOV strings*

We begin by investigating the interaction dynamics of STOV strings under continuously scanned dispersion. As shown in Fig. 1, the behavior depends critically on the charge distribution. For the STOVs with topological charge (TC) arrangement of (1,1,1), the three singularities do not simply compress temporally but execute a complex coupled motion in the spatiotemporal $(x, \xi)$ that resembles a 'vortex dance', seen in Fig. 1a,b. In contrast, introducing an opposite charge inbetween, with TC arrangement (1,-1,1), leads to an elongated singularity structure that evolves into filament-like lines, accompanied by lobe splitting and eventual annihilation (Fig. 1c). To make Fig. 1a,b quantitatively predictable, we first build a closed-form description for the (1,1,1) case.

To do so, we begin with an analytically convenient initial condition at group delay dispersion (GDD) = 0, where the input field can be written in a compact 'polynomial × Gaussian' form:

$$E(x, \tau, 0) = \left(\frac{\tau}{w_\tau} + i\frac{x}{w_x} + \rho\right)\left(\frac{\tau}{w_\tau} + i\frac{x}{w_x}\right)\left(\frac{\tau}{w_\tau} + i\frac{x}{w_x} - \rho\right) E_G(x, \tau) \quad (1)$$

In Eq. (1), $w_\tau$ and $w_x$ are the half-widths along the relative temporal coordinate $\tau = t - z/v_g$ and the spatial coordinate $x$, respectively. $E_G(x, \tau)$ is the Gaussian envelope, while $\rho$ sets the initial normalized spacing of the other singularities. As observed in Eq(1), the locations of all phase singularities are completely determined by the zeros of the polynomial prefactor. Consequently, once the dispersion-induced evolution of this prefactor is known, the singularity trajectories follow directly. To model the evolution under a programmed quadratic spectral phase, we use the paraxial propagation formalism for STOVs (see supplementary materials). In this picture, as dispersion increases, the wavepacket broadens along $\tau$ and accumulates a dispersion-induced Gouy-like phase. The corresponding dispersion-dependent width and phase given by $w_\tau = w_{\tau 0}\sqrt{1 + (GDD/w_{\tau 0}^2)^2}$, $\phi_\tau = \tan^{-1}(2GDD/w_{\tau 0}^2)$, where $k_0''$ is the group velocity dispersion (GVD). Substituting these dispersion-modified quantities into the propagated field gives the closed-form expression $E(x, \tau; GDD)$ (Method), and the corresponding singularity coordinates are then obtained by solving $E = 0$, which leads to the analytical singularity positions summarized in Eqs. (2) and (3).

$$(x_0, \tau_0) = (0,0) \quad (2)$$

$$(x_\pm, \tau_\pm) = \pm\left[w_x\left(\sqrt{\frac{M+R}{2}} - \sqrt{\frac{M-R}{2}}\tan\phi_\tau(GDD)\right), w_\tau(GDD)\left(\frac{\sqrt{\frac{M-R}{2}}}{\cos\phi_\tau(GDD)}\right)\right] \quad (3)$$

where the auxiliary quantities $R = \frac{3}{2}sin^2\phi_\tau(GDD) + \rho^2$ are defined as $I = \frac{3}{2}\sin\phi_\tau(GDD)\cos\phi_\tau(GDD)$, $M = \sqrt{R^2 + I^2}$. Eq. (3) reveals the spatial-temporal coupling nature of such STOVs. Although GDD is applied in the temporal domain, it drives a pronounced *spatial* drift of the singularities. As | GDD | is reduced toward the compression point and decreases toward zero, the outer singularities move monotonically toward the spatial axis $x$ seen in Fig. 1a, and their full spatiotemporal trajectories are plotted in Fig. 1b. Simultaneously, the temporal separation $\Delta\tau$ decreases. This concurrent reduction in both spatial $\Delta x$ and $\Delta\tau$ separations constitutes an effective **spatiotemporal attraction**, culminating in a minimum inter-singularity distance at the compression point (GDD=0). The minimum distance is governed by the parameter $\rho$; a larger $\rho$ corresponds to a weaker effective attraction, as seen in the increased minimum spacing. The interaction between two vortices can also be characterized by the slope $dx/dGDD$ and $d\tau/d(GDD)$ from Eq. (3) with respect to GDD, as shown in Fig. 1b.

For vortices with opposite topological charges, the evolution behavior changes qualitatively. As shown in Fig. 1c, the vortex–antivortex strings with TC (1,-1,1) does not remain as three well-separated point singularities as the dispersion is scanned. Instead, the existence of a opposite charge causes the singularity structure to stretch into curved filament-like lines, and the wavepacket can split into two lobes before undergoing annihilation. Because this process involves strong deformation and topological reconstruction of the field, we analyze it separately in the later section (details are provided in the supplementary S1).

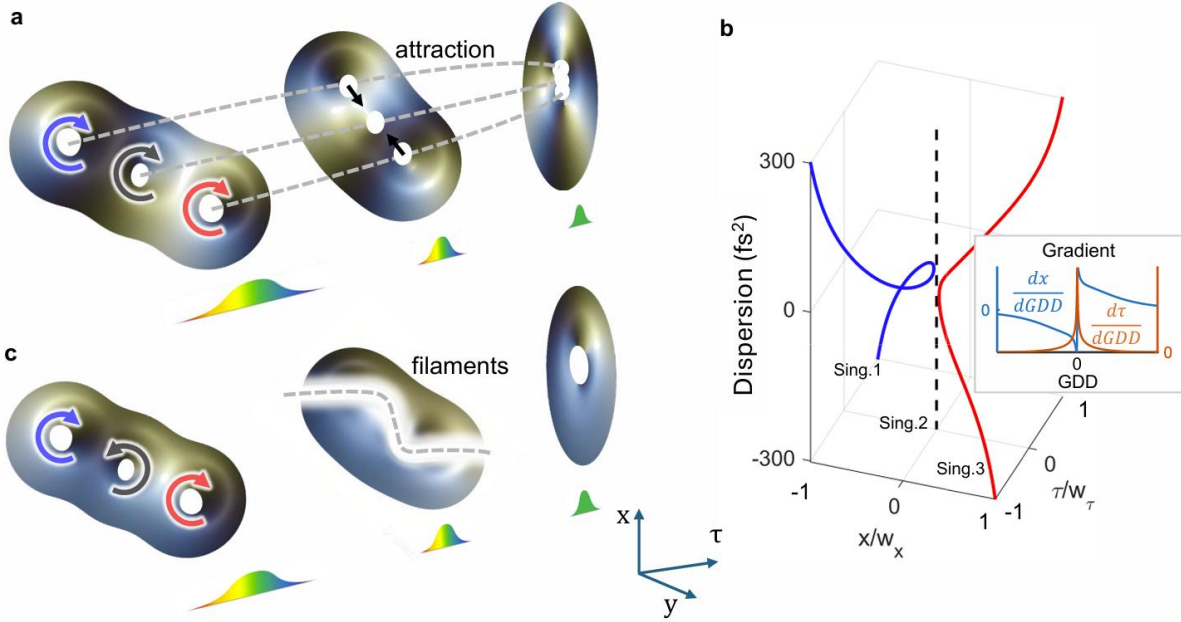

**Fig. 1. Typical images of interaction dynamics between transverse singularities.** **(a)** Three spatiotemporal vortex singularities with equal topological charges (1, 1, 1) gradually approach each other as temporal compression. **(b)** Plots of spatiotemporal positions of singularities with the (1, 1, 1) configuration of GDD. Gradient $dx/dGDD$ and $dt/dGDD$ of the spatiotemporal positions of singularities with the (1, 1, 1) configuration. **(c)** vortex–antivortex singularities (1, −1, 1) stretch to form singularity filaments.

To validate the theoretical framework, we built a time-frequency mapping pulse shaping system, using a spatial light modulator (SLM) to modulate the wave packet in the spatial-frequency domain, thereby embedding several spatiotemporal vortex singularities into one ultrafast wave packet. As shown in Fig. 2, experiment setup is as follows: a wave packet with about 50 fs generated by a Ti:sapphire mode-locked laser, passes through a pulse shaper composed of a grating, a cylindrical mirror, and a two-dimensional spatial light modulator (SLM). The plane of the SLM corresponds to the space-frequency $(x, \omega)$ plane of the wave packet, where the wave packet acquires a phase modulation $\varphi$ expressed as follows:

$$\varphi(x,\omega) = \tan^{-1}\left(\frac{x}{\omega + 0.5\omega_0}\right) + \tan^{-1}\left(\frac{x}{\omega}\right) + \tan^{-1}\left(\frac{x}{\omega - 0.5\omega_0}\right) + \beta\omega^2 \quad (4)$$

where ω and x are the coordinates of the SLM plane, and ω₀ denotes the initial frequency separation between vortices. A GDD phase is added to control the spatiotemporal distance. Subsequently, the wave packet passed through the cylindrical mirror and grating again to complete the temporal Fourier transform, while the focused far field enabled the spatial Fourier transform.

For wave packet characterization, it is feasible to exploit the Fourier transform properties of spatiotemporal pulses and reconstruct the full 3D spatiotemporal field by measuring the spatiospectral domain (28) (29). To accurately single-shot characterize the three-dimensional (3D) field of the wave packet, we propose a full interferometric retrieval of spatiotemporal tomography method (FIRST) for characterization of spatiotemporal wave packets (STWPs), as shown in Fig. 4. The FIRST method is characterized by acquiring complete spatiotemporal intensity and phase information of the 3D optical field $E(x, y, \tau)$ in a single shot and is well-suited for STWPs at the Fourier transform limit or with temporal dispersion. Moreover, the 3D information in both the spatial and temporal domains is retrieved via linear interference fringes, making it particularly suitable for single-shot 3D characterization of STWPs under low-light conditions. A grating combined with a narrowband filter was used to select the wave packet's frequencies, with each frequency component being spatially independent. Single-shot retrieval of the spatial intensity and spatial phase of all spectral components is achieved by analyzing the interference fringes. The spatiotemporal spectral phase information is obtained by sampling the spatial-spectral interferometric fringes at specific spatial positions using a spectrometer, which enables extremely high spectral phase calibration accuracy. This captures the complete 3D optical field information $E(x, y, \omega)$ and $E(x, y, \tau)$, allows for the observation of wave packets at arbitrary time scales in a single frame.

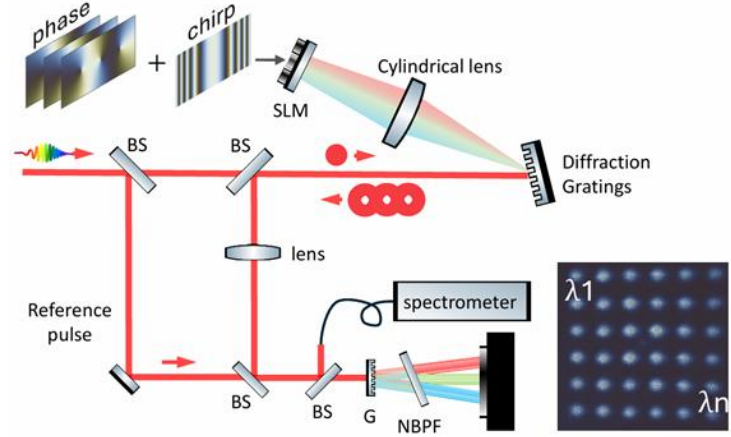

Fig. 2. Experimental setup. A linearly polarized Ti: sapphire mode-locked laser with wavelength λ = 800 nm is used. BS: beamsplitter, SLM: spatial light modulator, G: two-dimensional grating, NBPF: narrow bandpass filter.

Now, we discuss the singularities dynamics in spatiotemporal optical fields. To provide numerical simulation results for singularity prediction that are more consistent with the experimental setup, as well as intuitive visualization, we employed the spatiotemporal angular spectrum method (*28*) to simulate the complete optical field profiles of STWPs carrying multiple spatiotemporal vortex singularities propagating in dispersive media. Figs. 3a illustrates the propagation dynamics of STOVs under several dispersions, which includes three spatiotemporal vortex singularities with equal topological charges and mutual temporal delays. The temporal dispersion of the pulse governs the distance between spatiotemporal vortex singularities in the spatiotemporal domain. To investigate the dynamics of these singularities, we plot the light field maps in the x=0 plane and marked the singularities, as shown in Fig. 3.b1-b3. The singularity localization method is given in Supplementary Materials S2. Colored dot arrows indicate the positions and directions of the vortex singularities, where the color of each singularity is consistent with that in the subsequent figures. The black arrows indicate the movement direction of the singularities, which deviates from the direction of intuition along the temporal axis, this deviation underscores the spatiotemporally inseparable nature of spatiotemporal singularity dynamics. The singularities on both sides compress inward along the temporal axis; around -150 fs², Fig. 3. b2, they are gradually attracted to the central singularity, resulting in dynamic variations along the y-direction. Owing to the identical yet opposite attractive forces exerted by the singularities on both sides, the velocity and position of the central singularity remain unchanged. After completing the interactions in the main region, they return to their original trajectories and continue to extend toward infinity. Near the temporal Fourier transform limit, the STOV strings converge at the center and maintain a certain distance from each other, forming a dark field (Fig. 3(b3)). Interestingly, the vortices exhibit rapid and highly irregular tracer trajectories, which evokes parallels to the irregular Lagrangian trajectories during the collisions of fluid vortices (*30*), as both phenomena involve high-speed irregular internal trajectories when vortices approach or collide.

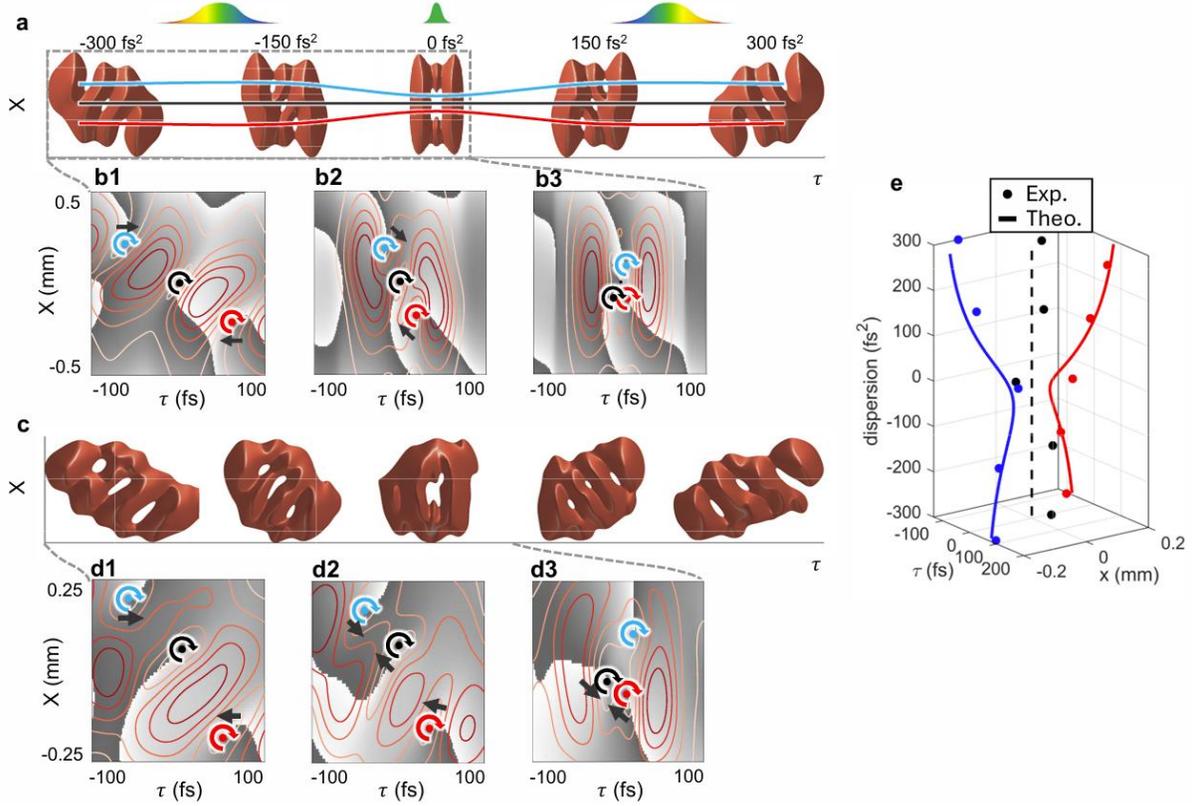

**Fig. 3. 3D Dispersive Dynamic Evolution of Spatiotemporal Optical Vortex with TC arrangement of (1 1 1). (a)** Dispersive dynamic evolution of the wave packet's intensity topological structure, where red surface represents iso intensity contours. **(b1-b3)** Displays the phase and modulus contour lines on 300 $fs^2$, 150 $fs^2$, 0 $fs^2$, reveal the variation of the singular points with different dispersions. The colored curved arrow represents vortices. The black arrow indicates the direction of motion of the singularity. As the pulse dispersion decreases, the blue singularity is attracted to and approaches the central black singularity. **(c)** Measurement results of the 1, 1, 1 wave packet experiment. **(d1-d3)** Experimental light field distributions at y=0, the black singularity first approaches the closer blue singularity and then moves toward the red one. **(e)** Comparison between experimental singularity tracking results and theoretical curves.

In the experimental results, as the wave packet undergoes temporal compression, the temporal distance between singularities decreases while their spatial distance also shrinks significantly. The isosurface of the wave packets are in excellent agreement with numerical simulations, as shown in Fig. 3c. The optical field slice at y=0, presented in Fig. 3d, shows that the three singularities emerge at the expected positions and gradually approach each other in the spatiotemporal domain. At GDD=0, the three singularities surround the vortex center. Notably, because the three singularities are not perfectly loaded with equidistant symmetry in the frequency domain in the experiment, the black singularity first approaches the closer blue singularity before moving toward the vicinity of the red one, which confirms the attractive dynamics of the singularities.

From Eq. 3, we note that this interaction process is related to the vortex distance ρ, which is the key parameter controlling the intensity of these dynamic behaviors. This can be controlled by the vortex distance $\omega_0$ in Eq. 4. We measured wave packets at three different values of $\omega_0$, namely 1/3, 2/3 and 1 times the spectral full width at half maximum $\omega_{FWHM}$; the singularity tracking results are shown in Fig. 4a, and the corresponding theoretical trajectories are presented in Fig. 4b, where the red, green, and blue curves represent the singularity distances in increasing order. Clearly, the blue curve is flatter, which is consistent with the theoretically predicted weaker spatiotemporal attractive effect.

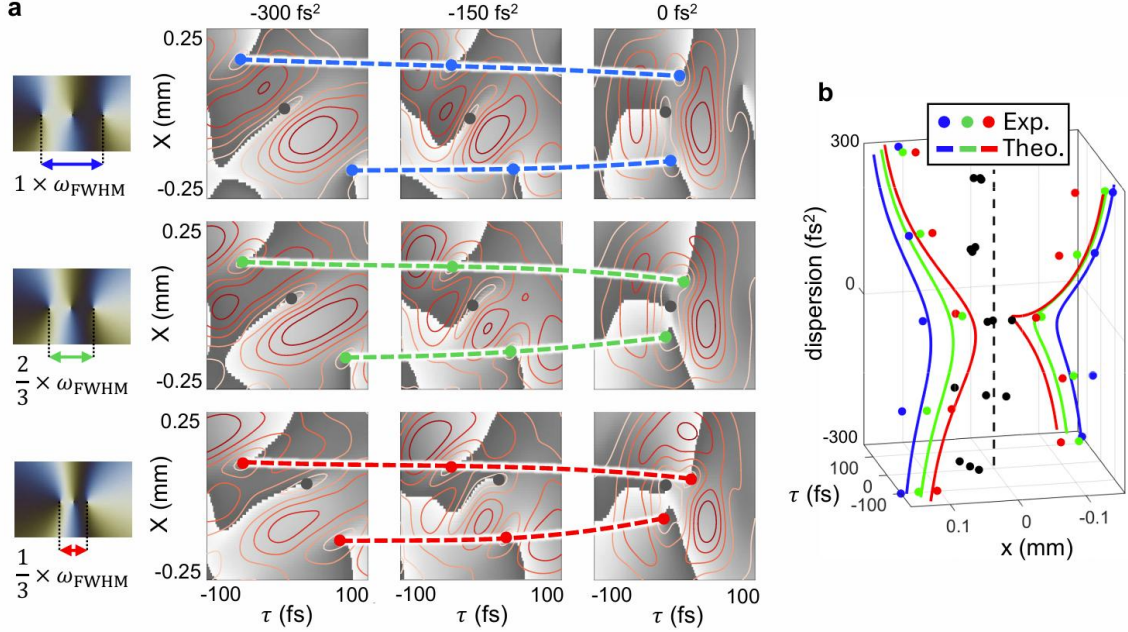

**Fig. 4. The dynamics of singularities at different singularity distances. (a)** Experiment results at three different values of ω0: 1/3, 2/3 and 1 times the spectral full width at half maximum $\omega_{FWHM}$. **(b)** The comparison between the experimental singularity tracking results and the theoretically predicted trajectories. The smaller the singularity distance, the more curved the curve becomes, indicating a stronger interaction.

### Annihilation of vortex and antivortex

Optical vortices and antivortices provide new degrees of freedom for modern optics and practical applications. (*25, 31-33*). We investigated the evolution of multiple vortices with opposite topological charges coexisting in an optical beam. When a wave packet contains both positive and negative vortex simultaneously, their interaction phenomena are more pronounced. We adjusted the topological charge of Vortex No. 2 to −1. Fig. 5 presents the numerical simulation and experimental results of the interaction dynamics of vortices with a TC (1, -1, 1) distribution, as well as the interaction process of positive and negative STOV. At large dispersion, the three vortex singularities with topological charges of 1, -1, and 1 are sequentially distributed along the temporal dimension. When the wave packet is compressed to 150 fs² (Fig. 5a), the vortex-antivortex begin to attract and stretch each other, forming filaments that grow along the direction perpendicular to the collision direction. At this stage, the wave packet splits into two lobes. Subsequently, the vortex lines rapidly reconstruct and annihilate (Fig. 5b3), leaving a spatiotemporal toroidal structure with L=1 remaining in the spatial domain. While the structure at this position is highly unstable, with further increases in dispersion, the toroidal spatiotemporal vortex ruptures again, splitting into two separate fragments in an S-shape. Contrary to Fig. 5b2, this splitting occurs in the opposite direction due to the reversal of dispersion.

Fig. 5c shows the experimental results for the (1, -1, 1) distribution. During the dispersion process, the positive and negative spatiotemporal vortex inevitably interact with each other, stretching into curved dark-field filaments. Near zero dispersion, the wave packet re-closes into a spatiotemporal donut with an annular energy distribution, leaving only one vortex singularity and two phase shifts. All experimental results are consistent with the simulation results and theoretical laws. Interestingly, the field we observed during the interaction of positive and negative STOV strings are similar to the physical picture of positive and negative fluid vortex pair interactions(*34*), which both exhibit three characteristic intensity distributions.

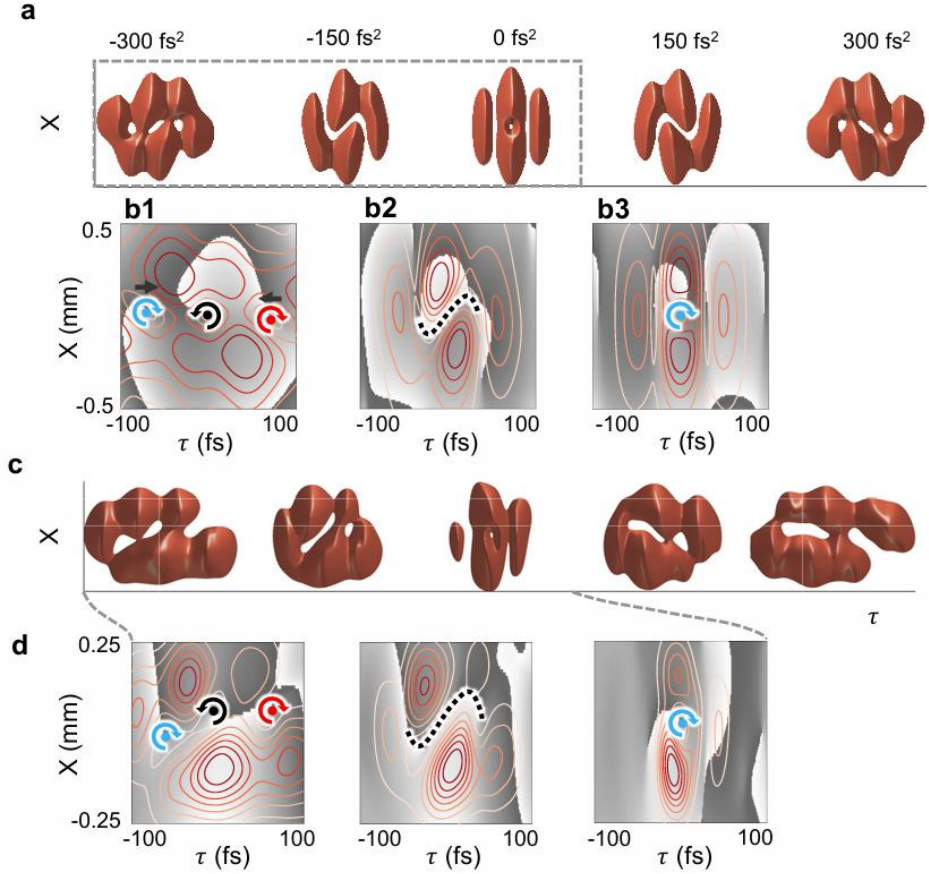

**Fig. 5. 3D Dispersive Dynamic Evolution and Optical Field Vectors of Spatiotemporal Optical Vortex with (1, -1, 1) Distribution.**
**(a)** The dispersive dynamic evolution of the wave packet. **(b1-b3)** Displays the phase and modulus contour lines at x=0 plane, revealing the variation of the singular points with different dispersions. The colored curved arrows represent the position and direction of the vortex singularities. **(c)** Experiment results of the (1, -1, 1) wave packet. **(d)** Experimental light field distributions at y=0.

## Discussion

Dynamical phenomena such as the attraction, collision, and annihilation of vortex singularities have been extensively studied in fluid dynamics. The generation and annihilation of vortex-antivortex also represent key topics in condensed matter physics (*35, 36*), we have observed these phenomena similarly in spatiotemporal optical fields. Notably, the study of interactions and dynamics between vortex singularities is a research hotspot in optics (*25, 33, 37, 38*); we believe our demonstration will draw new attention to novel perspectives on spatiotemporal singularity dynamics. The three characteristic phenomena we observed, independently distributed vortices, stretching into curved filaments, and annihilation between the singularities of positive and negative spatiotemporal vortex are highly similar to the three characteristic physical behaviors of positive and negative vortex pair interactions observed in aerodynamic fluid wakes (*39*). This cross-field similarity points to potential similarities in topological dynamical laws, whether they happen in optical or fluid media.

Our theory and experiments were performed by tuning the pulse chirp, Since temporal diffraction affects the toroidal vortex wave packet similarly to dispersion influences (*40*), this effect can be conveniently extended to the temporal diffraction regime. Meanwhile, we emphasize the necessity of precise measurement of such spatiotemporal structured wave packet during application; for instance, dispersion misalignment can lead to significant changes in the spatiotemporal structure of the wave packet. But these dynamic phenomena remain stable even when subjected to minor perturbations and aberrations in experiments. We hypothesize that this stability stems from the topological protection of spatiotemporal vortices, which will provide fundamental characteristic support for their applications in light-matter interaction, particle manipulation, and beyond.

In summary, in this work, we observed a series of intriguing phenomena within STOV, including the counterintuitive spatial-domain attraction of spatiotemporal vortex strings during temporal compression, the vortices and antivortices lead to stretching into filaments and annihilation. Meanwhile, we proposed the FIRST method, which enables single-shot 3D observation of complex STWPs with arbitrary time scales. Furthermore, the 3D information in both the spatial and temporal domains is retrieved via linear interference fringes, making this method suitable for single-shot 3D characterization of STWPs under arbitrary light intensity conditions. To the best of our knowledge, the intrinsic dynamics of interactions between transverse vortex singularities inherent to these optical STWPs are reported here for the first time, opening new avenues for research in the field of spatiotemporal electrodynamics.

We anticipate potential applications: for instance, the attraction of spatiotemporal vortices and singularity filaments could be used to manipulate accelerated charged particles in light-matter interactions, while the collision and annihilation phenomena may find applications in the field of optical encryption or communication. Given the universality of the underlying principle, We propose that singularity dynamics in the spatiotemporal domain constitute a universal phenomenon, applicable to various types of vortex fields—such as fluid vortices (*41*), quantum waves (*42*), electron waves, and acoustic waves (*14*). Related studies have shown that interaction processes can be investigated via the Gross–Pitaevskii equation for Bose-Einstein condensates (BECs) or multidimensional Schrödinger equations (*33, 37, 43*).

## Materials and Methods

The ultrashort pulse originated from the oscillator (800 nm central wavelength, 75-MHz repetition frequency, p polarization, 50-fs pulse duration) and entered the 4F spatial–spectral (x, ω) pulse shaper to generate the STOV. This setup included a blazed grating (1200 lines/mm, gold coating, with an incident angle of approximately 15°), a cylindrical lens (focal length of 100 mm), and a liquid crystal spatial light modulator (1920 × 1080 resolution, 6-μm pixel size, 60 Hz), on which the vortex phase is displayed.

The FIRST system includes a customized 2D grating (fused silica, 6 × 6 2D beam splitting, period of 21 μm, rotated by 10° along the optical axis), and a narrow bandpass filter (central wavelength of 830 nm and an FWHM of 0.5 nm) placed close to the 6 × 6 grating. A complementary metal-oxide semiconductor (CMOS) camera (resolution of 4096 × 4096, pixel size of 9 μm, and 16-bit monochromatic). The spectral resolution of the spectrometer is approximately 0.5 nm.

**Acknowledgments**

**Funding:**
This work was supported by the National Natural Science Foundation of China (62575295，12574368，12004403, 12074399, 12204500, 52327802 and 62227821);
Science and Technology Commission of Shanghai Municipality (25692112900);
Chinese Academy of Sciences (XDA25020105, CXJJ-23S015);
Ministry of Science and Technology (2021YFE0116700);

**Author contributions:**
Conceptualization: X. Y., P. Z. and X. X.

theoretical analysis and simulations: X. Y., X. G., J. R., J. F., M. H., and X. L.
experimental measurements: X. Y., P. Z., and Z. G.
analyzed the data: X. Y., X. Z., M.S., Q. Z., L. C., H. Z., and J. Z.

**Competing interests:** Authors declare that they have no competing interests.

**Data and materials availability:** The data that support the plots within this paper are available from the corresponding author upon reasonable request. Source data are provided with this paper.